\documentclass[preprint,12pt]{elsarticle}

\usepackage{amssymb}

\journal{Physics Letters A}

\begin{document}

\begin{frontmatter}



\title{Optimal control of population transfer in Markovian open quantum systems}


\author[AMSS,GUCAS]{Wei Cui}
\author[AMSS]{Zairong Xi\corref{cor}}\ead{zrxi@iss.ac.cn}
\cortext[cor]{Corresponding author at: Key Laboratory of Systems and
Control,  Academy of Mathematics and Systems Science, Chinese
Academy of Sciences, Beijing 100190, P.~R.~China}
\author[AMSS,GUCAS]{Yu Pan}
\address[AMSS]{Key Laboratory of Systems and Control,  Academy of Mathematics and Systems Science, Chinese Academy
of Sciences, Beijing 100190, P.~R.~China}
\address[GUCAS]{Graduate University of Chinese Academy of Sciences, Beijing
100039, P.~R.~China}

\begin{abstract}
There has long been interest to control the transfer of population
between specified quantum states. Recent work has 
optimized  the control law for closed system population transfer by
using a gradient ascent pulse engineering algorithm
\cite{Schonfeldt:09}. Here, a spin-boson model consisting of
two-level atoms which interact with the dissipative environment, is
investigated. With optimal control, the quantum system can invert
the populations of the quantum logic states. The temperature plays
an important role in controlling population transfer. At low
temperatures the control has active performance, while at high
temperatures it has less effect. We also analyze the decoherence
behavior of open quantum systems with optimal population transfer
control, and we find that these controls can prolong the coherence
time. We hope that active optimal control can help quantum
solid-state-based engineering.

\end{abstract}


\end{frontmatter}



\section{Introduction}
There exists a widespread belief that future technologies will
employ typical quantum features of microscopic systems, like
entanglement or superpositions of quantum states. The most prominent
examples are the rapid development of quantum information theory and
computation theory \cite*{Stolze:08,Bellac:06,Mermin:07}, including
 realizing high-speed quantum computation, and
 high-security quantum communication.
Both quantum computation and quantum information are the study of
information processing tasks that can be accomplished in quantum
mechanical systems.
In quantum physics, a system composed of two
or more quantum subsystems can be in a state that has no classical
counterpart. It is in a superposition of two or more quantum states
with a precise phase of the whole system. This property is quantum
coherence, which can  show entanglement. Control of quantum
coherence and entanglement is of great importance
\cite*{Wiseman:93,Wiseman:10,Rabitz:04,Rabitz:10,Balint:08,Facchi:05,Zhang1,Dong:09,Wu:06}.
The control of closed quantum systems is well established
\cite*{Tarn:83}. Efforts to extend these studies to open quantum
systems \cite{Breuer:02}, where the systems of interest interact
with their surrounding environments, are now underway. The aim of
dynamical control in open quantum systems is to suppress effects of
the environment in order to preserve the quantum properties,
including quantum coherence and entanglement, etc. The Zeno effect
has been used to achieve certain level of control on quantum
systems, inhibiting the decay of excited states due to many repeated
measurements. This form of quantum control has been studied in
several works, including \cite{Facchi:05,Wang:08,Zhou:09,Cao:10}.

 Superconducting
qubits \cite*{You:05} have also been studied as ways to control and
interact with naturally formed quantum two-level systems in
superconducting circuits. The two-level systems naturally occurring
in Josephson junctions constitute a major obstacle for the operation
of superconducting phase qubits. Since these two-level systems can
possess remarkably long decoherence times, References
\cite{Ashhab1,Ashhab2,Ashhab3} showed that such two-level systems
can themselves be used as qubits, allowing for a well controlled
initialization, universal sets of quantum gates, and readout. Thus,
a single current-biased Josephson junction can be considered as a
multi-qubit register. It can be coupled to other junctions to allow
the application of quantum gates to an arbitrary pair of qubits in
the system. These results \cite{Ashhab1,Ashhab2,Ashhab3} indicate an
alternative way to control qubits  coupled to naturally formed
quantum two-level systems, for improved superconducting quantum
information processing.  Indeed, these predictions have been found
experimentally in \cite{Neely:08}. More recently, reference
\cite{Burgarth:09} applies quantum control techniques to control a
large spin chain by only acting on two qubits at one of its ends,
thereby implementing universal quantum computation by a combination
of quantum gates on the latter and swap operations across the chain.
They \cite{Burgarth:09} show that the control sequences can be
computed and implemented efficiently. Moreover, they discuss the
application of these ideas to physical systems such as
superconducting qubits in which full control of long chains is
challenging.

%
Considerable experimental and theoretical attentions have been paid
in this field, especially using superconducting quantum circuits.
Reference \cite{Liu:05} analyzes the optical selection rules of the
microwave-assisted transitions in a flux qubit superconducting
quantum circuit. They \cite{Liu:05} show  that the parities of the
states relevant to the superconducting phase in the charge qubit can
be controlled via the external magnetic flux. For certain values of
the flux,  the selection rules are the same as the ones for the
electric-dipole transitions in usual atoms. In other cases,  the
symmetry of the potential of the artificial ``atom" is broken, a
so-called $\Delta$-type ``cyclic" three-level atom is formed, where
one- and two-photon processes can coexist.  They also study how the
population of the three states can be selectively transferred by
adiabatically controlling the electromagnetic field pulses.
Different from $\Delta$-type atoms, the adiabatic population
transfer in that three-level atom can be controlled not only by the
amplitudes but also controlled by the phases of the pluses. Thus,
they achieved a pulse-phase-sensitive adiabatic manipulation of
quantum states in this three-level artificial atom. Usually, only
the amplitude was considered for adiabatic control. This example of
control on quantum circuits has been recently studied experimentally
\cite{Deppe:08}. In \cite*{Wei:08}, a novel approach was proposed to
coherently transfer populations between selected quantum states in
one and two qubit systems by using controllable Stark-chirped rapid
adiabatic passages (SCRAPS) (see Fig.1). These time-insensitive
evolution transfers, assisted by easily implementable single-qubit
phase-shift operations, could offer an attractive approach to
implement high-fidelity single-qubit NOT operations for quantum
computing. Specifically, this proposal could be conveniently
demonstrated by existing Josephson phase qubits.

\begin{figure}
\begin{center}
\includegraphics[width=0.9\textwidth,bb=2 2 810 200]{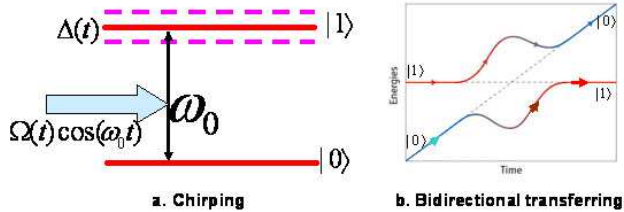}
\caption{(Color online)a. Schematic view of inducing population
transfer from the ground state to the excited state\cite{Wei:08}. b.
Schematic view of single-qubit logic gate\cite{Wei:08}.}
\label{schematic}
\end{center}
\end{figure}

However, a quantum system can never be isolated from the surrounding
environment completely  \cite*{Breuer:02,Weiss:99}. As a result, a
randomization of the phase of the quantum system takes place, and
the initial quantum state ends up in a classical state.
 Thus, it is an important subject to analyze
the quantum decay induced by the unavoidable interaction with the
environment. Population transfer control is an important problem
with application to various quantum systems, and many results are
obtained  with closed-loop learning control and quantum optimal
control for large transfer time, respectively. The SCRAP-based
quantum gates proposed in
 \cite*{Wei:08}  evolve in the closed quantum system, in which the
surrounding environment effect was not considered. Recently, J. H.
Sch\"{o}nfeldt, J. Twamley and S. Rebi\'{c} considered the closed
system population transfer control in stark-shift-chirped rapid
adiabatic-passage (SCRAP) technology \cite{Schonfeldt:09}. Their
main result is that by the gradient ascent pulse engineering
algorithm the average fidelity of population transfer over a wide
range of detunings for both the ground to excited state detuning and
the ground to target state detuning can be improved. In this paper,
we make use of general optimal control theory  to transfer the
quantum population for quantum computing in Markovian open,
dissipative quantum systems.

 The paper
is organized as follows. 
In Sec. II we introduce the controlled spin-boson model. The
Lindblad type master equation for driven open quantum systems with
the noise and dissipation effects are presented in this section. In
Sec. III, we analyze the optimal control of quantum system dynamics.
Both theoretical analysis and numerical demonstration are presented
in this section. Conclusions and prospective views are given in Sec.
IV.

\section{Master equation}
A quantum two-level system coupled to an environment is always
modeled by a spin degree of freedom in a magnetic field coupled
linearly to an oscillator bath with Hamiltonian \cite*{Makhlin:01}
\begin{equation}
H=H_{C}+H_{B}+H_{I}.\label{Hamiltonian}
 \end{equation}
Here the oscillator bath is described by
$H_B=\sum_k\hbar\omega_ka_k^{\dag}a_k$; the spin's observable
$\sigma_z$ is coupled with the bath ``force" operator $X=\sum_k(g_k
a_k^{\dag}+g_k^{*}a_k)$. In the following, the Planck constant
$\hbar$ is assumed to be $1$. Let the controlled part
 \begin{eqnarray}
\label{control}
H_{C}&=&\frac{1}{2}B_z\sigma_z+\frac{1}{2}B_x\sigma_x\nonumber\\&=&\frac{1}{2}\Delta
E(\cos\eta\sigma_z+\sin\eta\sigma_x),
 \end{eqnarray}
where the mixing angle $\eta\equiv\arctan(B_x/B_z)$ determines the
direction of the effective magnetic field in the $x-z$ plane, and
the energy splitting between the eigenstates is $\Delta
E=\sqrt{B_x^2+B_z^2}$. $B_x$, $B_z$ are the external controls. For
the free system Hamiltonian $H_0=\frac{\omega_0}{2}\sigma_z$ with
$\omega_0=1$ as the norm unit throughout this paper, and $H_0$ can
be contained in $H_{C}$. Usually, identifying the nature of
interactions in a quantum system is essential in understanding it.
Acquiring information on the Hamiltonian can be difficult for
many-body systems because it generally requires access to all parts
of the system. Reference \cite{Burgarth:092} showed that if the
coupling topology is known, the Hamiltonian identification is indeed
possible indirectly even though only a small gateway to the system
is used. Surprisingly, even a degenerate Hamiltonian can be
estimated by applying an extra field to the gateway.  This
information can then be used for achieving a better control of the
quantum system.

Under the adiabatic approximation, it is natural to describe the
evolution of the system in the eigenbasis of $H_{C}$. From
Eq.(\ref{control}), the eigenvalues are $\pm\frac{1}{2}\Delta E$ and
the corresponding instantaneous eigenstates are
 \begin{equation}
|\lambda_{+}(t)\rangle=\cos\frac{\eta}{2}|0\rangle+\sin\frac{\eta}{2}|1\rangle,
 \end{equation}
 \begin{equation}
|\lambda_{-}(t)\rangle=-\sin\frac{\eta}{2}|0\rangle+\cos\frac{\eta}{2}|1\rangle.
 \end{equation}

The goal of the theoretical treatment is to analyze decoherence and
the controlled population transfer, which is given by the elements
of the reduced density matrix, defined as
$\rho_{S}(t)=tr_{B}[\rho_{tot}(t)],$ where $\rho_{tot}$ is the total
density matrix for both the system and the environment, $tr_{B}$ the
partial trace taken over the environment. The effect of the
environment on the dynamics of the system can be seen as a interplay
between the dissipation and fluctuation phenomena.

The simplest quantum system is a two-level system, whose Hilbert
space is spanned by two states, an excited state $|e\rangle$ and a
ground state $|g\rangle$. The Hilbert space of such a system is
equivalent to that of a spin-$\frac{1}{2}$ system. The corresponding
Pauli operators are $\sigma_1=|e\rangle\langle g|+|g\rangle\langle
e|, \sigma_2=-i|e\rangle\langle g|+i|g\rangle\langle e|,
\sigma_3=|e\rangle\langle e|-|g\rangle\langle g|,$
satisfying the commutation relations$ [\sigma_i,
\sigma_j]=2i\varepsilon_{ijk}\sigma_k,$ and the anticommutation
relations $ \{\sigma_i, \sigma_j\}=2\delta_{ij}$.
 Then the controlled master equation of two-level system has the following Lindblad form
 \cite{Breuer:02} ($\rho$ instead of $\rho_S$ for simplicity),
\begin{eqnarray}
\frac{d\rho(t)}{dt}=-i[H_{C},
\rho]+\left.\frac{\gamma_0(N+1)}{2}\right\{2\sigma
\rho\sigma^{+}-\sigma^{+}\sigma\rho\nonumber\\-\rho\sigma^{+}\sigma\left\}+
\left.\frac{\gamma_0N}{2}\right\{2\sigma^{+}\rho\sigma-\sigma\sigma^{+}\rho
-\rho\sigma\sigma^{+}\right\}, \label{master equation}
\end{eqnarray}
where $N=1/(e^{\frac{\hbar \omega_0}{K_BT}}-1)$ denotes the mean
number of quanta in a mode with frequency $\omega_0$ of the thermal
reservoir, $\gamma_0$ is the spontaneous emission rate. The matrix
elements $\rho_{00}=p_g(t)$ and $\rho_{11}=p_e(t)$ are the
population of the ground and excited state levels, respectively. The
off-diagonals $\rho_{01}(t)=\rho_{10}^{*}(t)$ are the coherences.
Let $x(t)=Tr[\sigma\rho(t)],$ which is called as Bloch vector and
defined by
\begin{equation}
\label{Bloch vector}\left\{
 \begin{array}{rcl}
x_1(t)&\equiv&\rho_{01}(t)+\rho_{10}(t),\\
x_2(t)&\equiv&i(\rho_{10}(t)-\rho_{01}(t)),\\
x_3(t)&\equiv&\rho_{00}(t)-\rho_{11}(t) .
\end{array}\right.
 \end{equation}
Then the Bloch vector form of (\ref{master equation}) is
\begin{equation}
\label{lindblad components}
 \begin{array}{rcl}
\dot{x_1}(t)&=&-\frac{2N+1}{2}\gamma_0x_1(t)+B_z x_2(t),\\
\dot{x_2}(t)&=&-B_z x_1(t)-\frac{2N+1}{2}\gamma_0x_2(t)+B_x x_3(t),\\
\dot{x_3}(t)&=&-B_x x_2(t)-(2N+1)\gamma_0x_3(t)-\gamma_0,
\end{array}
 \end{equation}
which can be written compactly as

\begin{equation}
\dot{x}(t)=A(t)x(t)+B(t)
 \end{equation}
where
\[
A(t)=\left(\begin{array}{ccc}
-\frac{2N+1}{2}\gamma_0&B_z&0\\
-B_z&-\frac{2N+1}{2}\gamma_0&B_x\\
0&-B_x&-(2N+1)\gamma_0
\end{array}\right)
\]
and
\[
B(t)=\left(\begin{array}{ccc}
0\\
0\\
-\gamma_0
\end{array}\right).
\]


\section{Optimal control of population transfer}
\subsection{Optimal control formalism}
Recent work \cite{Schonfeldt:09} has optimized  the control law for
closed system population transfer by using a gradient ascent pulse
engineering algorithm. They shows that the optimized pulses perform
at a higher fidelity than the standard Gaussian pulses for a wide
range of detunings (i.e., large inhomogeneous broadening). The
theory of optimal control was introduced in the theory of automatic
control in the 1960s for electrical engineering applications.
Bellman and Pontryagin, the two famous scientists, paved the way in
this field, respectively.  Prof. Belavkin introduced it to quantum
mechanical in 1983 \cite{Belavkin:83}. In quantum chemical, quantum
optimal control made great success
\cite{Rice,Gordon,Brumer,Rabitz:10,Alessandro}.
In the optimal control framework, one starts by defining a cost
functional which has the function of optimality criteria. Moreover,
the cost functional will vary from one experimental trial to
another, and must be thought of as a random variable depending on
the measurement output. The strategy is then to minimize this cost
functional while satisfying the constraints of the underlying
dynamic equations governing the evolution of quantum states, e.g.,
the master equation and the initial state condition. The calculation
of the necessary optimality conditions for this optimization problem
results in a system of coupled equations to be solved. For detail
one can refer to the papers
\cite{Rabitz:10,Balint:08,Rebentrost:09,Jirari:06}. Obviously, the
evolution of the state variable $x(t)$ governed by the master
equation (\ref{lindblad components}) depends not only on the initial
state $x_0$ but also on the choice of the time-dependent control
variable $u(t)=(B_x(t),~B_z(t))^T$. In this section, we are going to
control population transfer and suppress the unexpected effect of
decoherence by optimal control technique that wants to force the
system evolving along some prescribed cohering trajectories. For
some target state  $x^0(t)$, let the cost functional as
\begin{equation}
J[u(t)]=\Psi[x(t_f),x^0(t_f)]+\int_{t_o}^{t_f}L(x(t),x^0(t_f),u(t))dt,
 \end{equation}
where the functional $\Psi[x(t_f),x^0(t_f)]$ represents some
distance between the system and objects at final time and the
functional $\int_{t_o}^{t_f}L(x(t),x^0(t_f),u(t))$ accounts for the
transient response with $L(x(t),x^0(t_f),u(t))\geq 0$. The optimal
control problem considered in this paper is to minimize the cost
functional $J[u(t)]$ with dynamical constraints (\ref{lindblad
components}) and initial state constraint $x(0)$. Using the
Pontryagin's maximum principle, the optimal solution to this problem
is characterized by the so-called Hamilton-Jacobi-Bellman(HJB)
equation. We use this method to solve the population transfer and
decoherence in the following.

\begin{figure*}
\centerline{\scalebox{0.8}[0.6]{\includegraphics{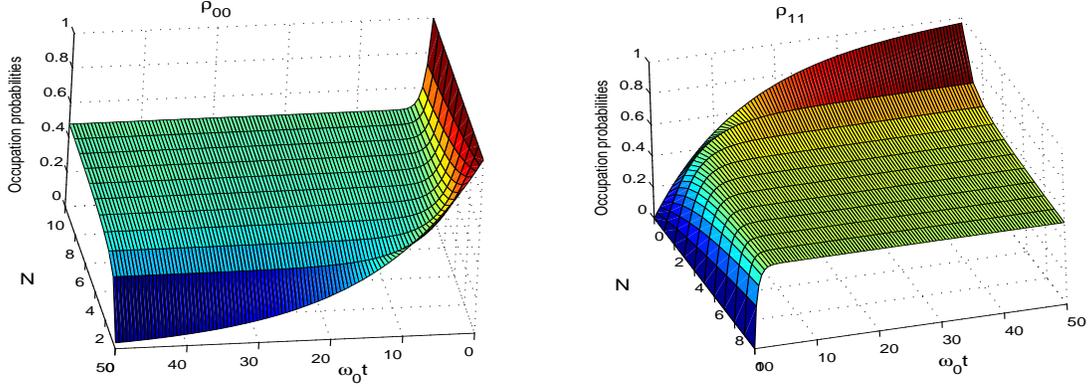}}}
\caption{(Color online)Time evolution of occupation probabilities
for $N\in[0,10]$ without control.}
\end{figure*}

\subsection{Evolution without control}
At first we consider the free evolution of the system (\ref{master
equation}), where the external control field $B_x=0,~B_z=\omega_0$.
Then the free evolution is
\begin{equation}
\label{free evolution}\left\{
 \begin{array}{rcl}
\dot{x_1}(t)&=&-\frac{2N+1}{2}\gamma_0x_1(t)+\omega_0 x_2(t),\\
\dot{x_2}(t)&=&-\omega_0 x_1(t)-\frac{2N+1}{2}\gamma_0x_2(t),\\
\dot{x_3}(t)&=&-(2N+1)\gamma_0x_3(t)-\gamma_0,
\end{array}\right.
 \end{equation}
whose solution is
\begin{equation}
x_1(t)=e^{-\frac{2N+1}{2}\gamma_0t}[x_2(0)\sin(\omega_0t)+x_1(0)\cos(\omega_0t)],
 \end{equation}
\begin{equation}
x_2(t)=e^{-\frac{2N+1}{2}\gamma_0t}[x_2(0)\cos(\omega_0t)-x_1(0)\sin(\omega_0t)],
 \end{equation}
\begin{equation}
x_3(t)=e^{-(2N+1)\gamma_0t}\left(\frac{1}{2N+1}+x_3(0)\right)-\frac{1}{2N+1}.
 \end{equation}
We observe that the population-component $x_3$ of the Bloch vector
decays exponentially with rate $-(2N+1)\gamma_0$, while the
coherence $x_{1,2}$ decay with rate $-(2N+1)\gamma_0/2$. The
stationary solution is
\begin{equation}
x_1^s=x_2^s=0,~~x_3^s=-\frac{1}{2N+1},
 \end{equation}
and the populations of the lower and upper level are found to be
\begin{eqnarray}
p_g(t)\equiv\rho_{00}(t)&=&\frac{1}{2}(1+x_3(t))\\\nonumber
&=&\frac{1}{2}e^{-(2N+1)\gamma_0t}\left[\frac{1}{2N+1}+2\rho_{00}(0)-1\right]+\frac{1}{2}\left(1-\frac{1}{2N+1}\right),
 \end{eqnarray}

\begin{eqnarray}
p_e(t)\equiv\rho_{11}(t)&=&\frac{1}{2}(1-x_3(t))\\\nonumber&=&-\frac{1}{2}e^{-(2N+1)\gamma_0t}\left[\frac{1}{2N+1}+2\rho_{11}(0)-1\right]+\frac{1}{2}\left(1+\frac{1}{2N+1}\right),
 \end{eqnarray}
with the stationary populations $\frac{1}{2}(1\pm\frac{1}{2N+1})$,
respectively. As $k_BT\rightarrow0$ the mean number of quanta
$N=1/(e^{\frac{\hbar \omega_0}{K_BT}}-1)\rightarrow0$, which means
that $\rho_{00}(t\rightarrow\infty)\rightarrow0$ and
$\rho_{11}(t\rightarrow\infty)\rightarrow1$. This is a population
transfer process. The elementary logic gates in quantum computing
networks can be implemented by this transfer. However, when $N>0$
$\rho_{00}(t\rightarrow\infty)=\frac{1}{2}(1-\frac{1}{2N+1})>0$, and
$\rho_{11}(t\rightarrow\infty)=\frac{1}{2}(1+\frac{1}{2N+1})<1$.
Furthermore, as $k_BT\rightarrow\infty$ the stationary populations
are
$\rho_{00}^S(N\rightarrow\infty)=\rho_{11}^S(N\rightarrow\infty)=\frac{1}{2}$,
which fails to transfer the populations. The evolution of
populations $\rho_{00}=\frac{1}{2}(1+x_3(t))$ (left) and
$\rho_{11}=\frac{1}{2}(1-x_3(t))$(right) is plotted for $N\in[0,10]$
in Fig.2, where the dissipation constant $\gamma_0$ is set as $0.1$,
and the the initial state
$x(0)=(\frac{\sqrt{2}}{2},\frac{\sqrt{2}}{2},1)$. In the following,
choosing the trajectory with $N=0$ as the target we will use the
optimal control technology to make the single qubit rotation that
completely inverts the populations of the quantum states.

 \begin{figure}
\centerline{\scalebox{0.9}[0.9]{\includegraphics{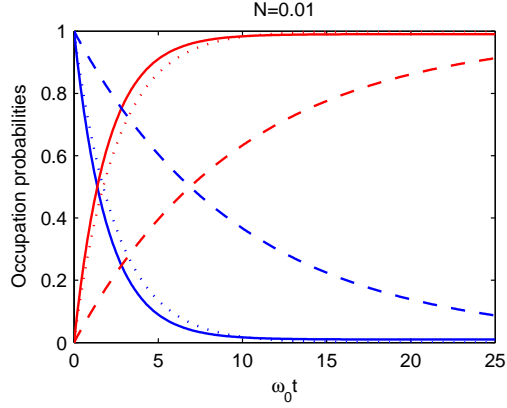}}}
\caption{(Color online)Time evolution of occupation probabilities
$\rho_{00}$(red) and $\rho_{11}$(blue), with system evolution
without control (the dashed line) and control target trajectory (the
dotted line).}
\end{figure}

\begin{figure}
\centerline{\scalebox{0.9}[0.9]{\includegraphics{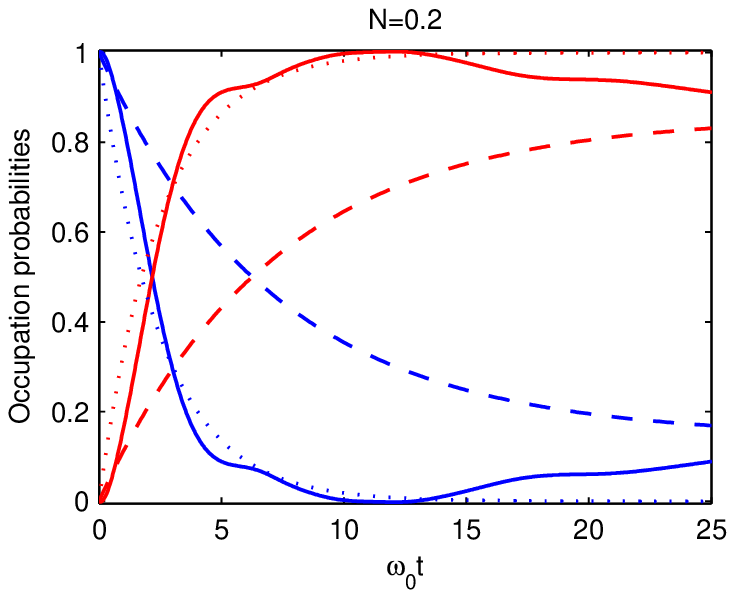}}}
\caption{(Color online)Time evolution of occupation probabilities
$\rho_{00}$(red) and $\rho_{11}$(blue), with system evolution
without control (the dashed line) and control target trajectory (the
dotted line).}
\end{figure}

\begin{figure}
\centerline{\scalebox{0.9}[0.9]{\includegraphics{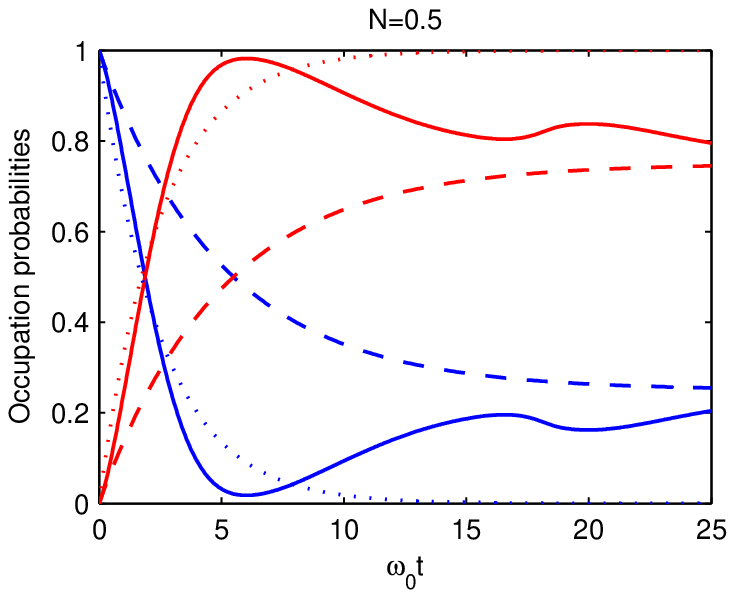}}}
\caption{(Color online)Time evolution of occupation probabilities
$\rho_{00}$(red) and $\rho_{11}$(blue), with system evolution
without control (the dashed line) and control target trajectory (the
dotted line).}
\end{figure}
 \subsection{Controlled population transfer}
Precise design of switching on/off logic gates of quantum systems is
a main problem in quantum computer research. In the following we
consider the optimal control of population transfer in a dissipative
open quantum system. For simplicity let the cost functional be:
 \begin{equation}
J[u(t)]=\int_{t_{0}}^{t_{f}}[(x(t)-x^0(t))^2+\theta u^{T}(t)u(t)]dt
 \end{equation}
where $\theta > 0$ is a weighting factor used to achieve a balance
between the tracking precision and the control constraints.
$u(t)=(B_x(t),~B_z(t))^T$ is the external control. $x^0(t)$ is the
target trajectory
\[
x^0(t)=\left(\begin{array}{c}
e^{-\frac{\gamma_0}{2}t}(x_2(0)\sin(B_zt)+x_1(0)\cos(B_zt))\\
e^{-\frac{\gamma_0}{2}t}(x_2(0)\cos(B_zt)-x_1(0)\sin(B_zt))\\
e^{-\gamma_0t}(x_3(0)+1)-1
\end{array}\right),
\]
which is the solution of the master equation (\ref{master equation})
in the case of $N=0$ without control, i.e.
$\frac{d\rho}{dt}=-i[\frac{\omega_0}{2}\sigma_z,
\rho]+\frac{\gamma_0}{2}\{2\sigma\rho\sigma^{+}-\sigma^{+}\sigma\rho-\rho\sigma^{+}\sigma\}$.
The dissipator of the equation describes spontaneous emission
process (rate $\gamma_0$) without thermally induced emission and
absorption process ($N=0$, or $T=0$). That was the reason we chose
it as the target trajectory. We will choose it as the tracking
object in both optimal population control in the following.

 The corresponding Hamiltonian function is
\begin{eqnarray*}
&&\mathcal{H}(x(t),u(t),\psi(t),t)\\&=&[(x(t)-x^0(t))^2+\theta
u^{T}(t)u(t)]+\lambda(t)^{T}[A(t)x(t)+B(t)]\\
&=&[(x_1(t)-x_1^0(t))^2+(x_2(t)-x_2^0(t))^2+(x_3(t)-x_3^0(t))^2
+\theta(B_x^2(t)+B_y^2(t))]\\&&+\lambda_1(t)[-\frac{2N+1}{2}\gamma_0x_1(t)+B_zx_2(t)]+\lambda_2(t)
[-B_zx_1(t)-\frac{2N+1}{2}\gamma_0x_2(t)\\&&
+B_xx_3(t)]+\lambda_3(t)[-B_xx_2(t)-(2N+1)\gamma_0x_3(t)-\gamma_0],
\end{eqnarray*}
 where $\lambda(t)=(\lambda_1(t),\lambda_2(t),\lambda_3(t))^{T}$ is the so-called Lagrange
 multiplier.
 The optimal solution can be solved by the following
 differential equation with two-sided boundary values,
\begin{equation}
\left\{  \begin{array}{rcl}
 \dot{x}(t)&=&\frac{\partial \mathcal{H}}{\partial
 \lambda}=A(t)x(t)+B(t),\\
 \dot{\lambda}(t)&=&-\frac{\partial \mathcal{H}}{\partial
 x}=-2[x(t)-x^0(t)]-A(t)^T\lambda(t),
 \\x(0)&=&x_0,\\
 \lambda(t_f)&=&0
 \end{array}\right.
 \end{equation}

\begin{equation}
 \begin{array}{rcl}
&&\Longrightarrow\\
&&\left\{  \begin{array}{rcl}
\dot{x_1}&=&-\frac{2N+1}{2}\gamma_0x_1+B_z x_2,\\
\dot{x_2}&=&-B_z x_1-\frac{2N+1}{2}\gamma_0x_2+B_xx_3,\\
\dot{x_3}&=&-B_xx_2-(2N+1)\gamma_0x_3-\gamma_0,\\
 \dot{\lambda_1}&=&-2(x_1-x_1^0)+\frac{2N+1}{2}\gamma_0\lambda_1+B_z\lambda_2,\\
 \dot{\lambda_2}&=&-2(x_2-x_2^0)-B_z\lambda_1+\frac{2N+1}{2}\gamma_0\lambda_2+B_x\lambda_3,\\
 \dot{\lambda_3}&=&-2(x_3-x_3^0)-B_x\lambda_2+(2N+1)\gamma_0\lambda_3,\\
 &&x(0)=x_0,~~~ \lambda(t_f)=0.
 \end{array}\right.\end{array}
 \end{equation}
 Together with
 \begin{equation}
 \left.\frac{\partial \mathcal{H}}{\partial u}\right|_*=\frac{\partial \mathcal{H}(x^*(t),u^*(t),\psi(t),t)}{\partial
 u}=0,
 \end{equation} we have
 \begin{equation}
\left\{  \begin{array}{rcl}
 B_x(t)&=&\frac{1}{2\theta}\{\lambda_3x_2-\lambda_2x_3\},\\
 B_z(t)&=&\frac{1}{2\theta}\{\lambda_2x_1-\lambda_1x_2\}.
\end{array}\right.
 \end{equation}
In general, no analytic solution of the complicated nonlinear
equations exists. The numerical demonstration to this problem is
considered in the following.

\begin{figure}
\centerline{\scalebox{0.9}[0.9]{\includegraphics{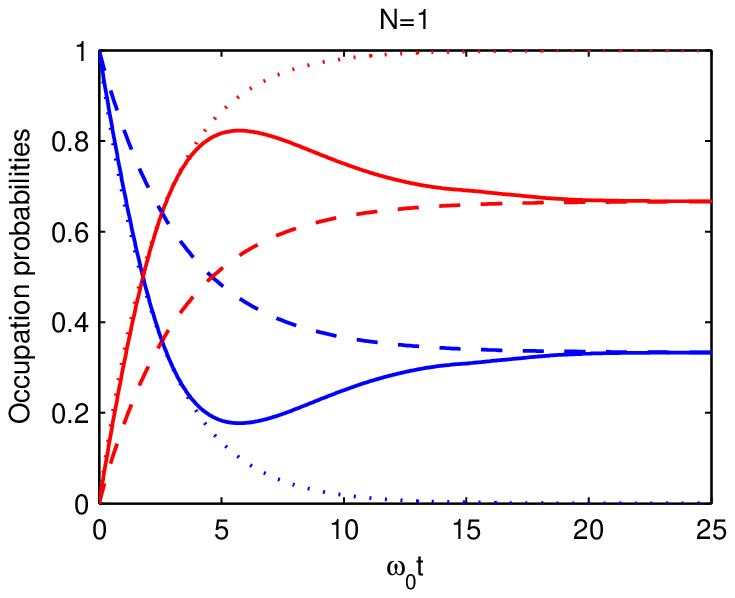}}}
\caption{(Color online)Time evolution of occupation probabilities
$\rho_{00}$(red) and $\rho_{11}$(blue), with system evolution
without control (the dashed line) and control target trajectory (the
dotted line).}
\end{figure}

\begin{figure}
\centerline{\scalebox{0.9}[0.9]{\includegraphics{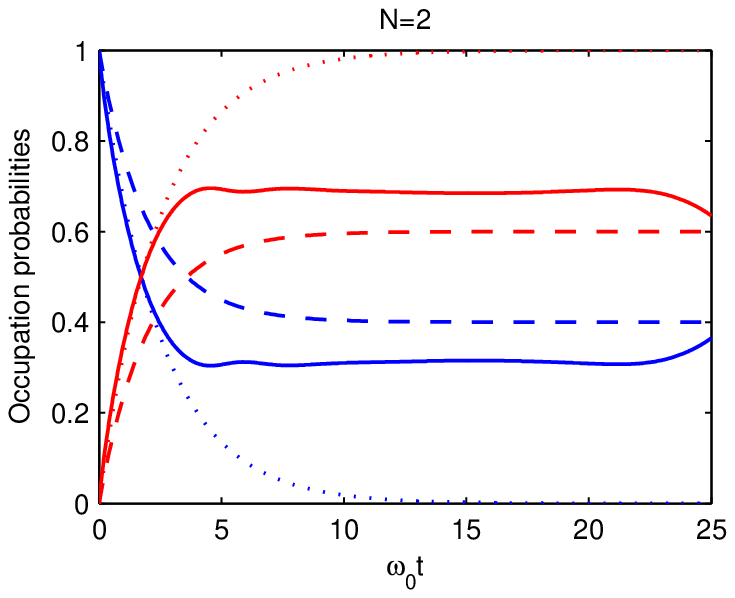}}}
\caption{(Color online)Time evolution of occupation probabilities
$\rho_{00}$(red) and $\rho_{11}$(blue), with system evolution
without control (the dashed line) and control target trajectory (the
dotted line).}
\end{figure}

\begin{figure}
\centerline{\scalebox{0.9}[0.9]{\includegraphics{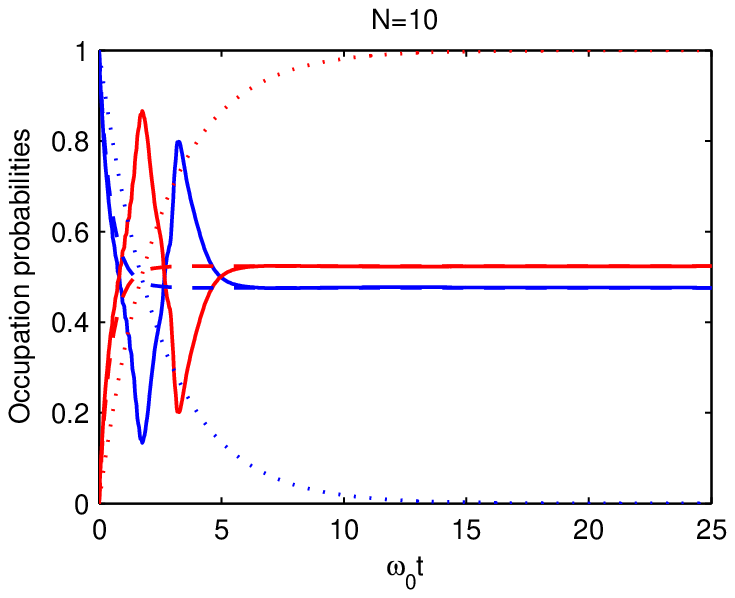}}}
\caption{(Color online)Time evolution of occupation probabilities
$\rho_{00}$(red) and $\rho_{11}$(blue), with system evolution
without control (the dashed line) and control target trajectory (the
dotted line).}
\end{figure}

\subsection{Numerical demonstration and discussions}
As example we show in Fig.(3-8) the populations as a function of
time for the open quantum system with six different values of
$N$(temperature). In our simulations, the system parameters are
chosen as: $x(0)=(\sqrt{2},\sqrt{2},1)$, dissipation constant
$\gamma_0=0.1$, and system frequency $\omega_0=1$ as the norm unit.
We consider the two-mode control, physically, $B_z(t)$
time-dependent Stark shift and $B_x(t)$ external resonant control
field acting on the open system. In our simulations, the time
evolutions of the lower level population
$p_g(t)\equiv\rho_{00}(t)=\frac{1}{2}(1+x_3(t))$ are colored red and
the time evolutions of the upper level population
$p_e(t)\equiv\rho_{11}(t)=\frac{1}{2}(1-x_3(t))$ are colored blue,
and the free evolutions are plotted with dashed line, the control
target trajectory with dotted line, optimal populations control with
solid line.

 More precisely, we study the temperature as a key
factor in population transfer control. In fact,
$N=1/(e^{\frac{\hbar\omega_0}{k_BT}}-1)$, which means that
$T=\hbar\omega_0/(k_B\ln(1+\frac{1}{N}))$, where
$k_B=1.380662\times10^{-23}J\cdot K^{-1}$ is the
Boltzmann constant 
and $\hbar=1.0545887\times10^{-34}J\cdot s$ is the reduced Planck
constant.
When $N$ is from $0$ to $10$, the temperature changes significantly,
$(0\sim8.0182)\times10^{-11}\omega_0 ~(K)$. The smaller the $N$ the
lower the temperature, the larger the $N$ the higher the
temperature. Looking at Figs. (3-8)(where we set $N$ by
$0.01,~0.2,~0.5,~1,~2,~10$, respectively), we can approximately say
that when the temperature is low, for example, $N<0.5~
(T=0.6956\times10^{-11}\omega_0 ~(K))$, we can achieve very good
population transfer fidelity. And the lower the better. Increasing
$N$, smaller population transfer were observed in Fig.(6) ($N=1$)
and Fig.(7) ($N=2$). This is because in higher temperature regime
the energy level spacing is decreased, which results in low transfer
rate. When $N=10$, the population is wild oscillations making the
optimal control failure. It means that the environment induced
fluctuations is large enough, so that the control field is
negligible comparing with the high-frequency harmonic oscillators of
the reservoir. From Figs. (3-8) we can sum up three rules. (i) The
external control field played an important role in population
transfer for quantum computing. (ii) The smaller the $N$
(temperature) the better the control performance, especially when
$N<0.2$ the completely population transfer can be achieved. (iii)
When temperature is high enough the transfer is uncontrollable.

\subsection{Decoherence behavior}
In the following we make efforts to study the decoherence behavior
of the open quantum systems with optimal population transfer
control, briefly. The persistence of quantum coherence is relied on
in quantum computer, quantum cryptography, quantum teleportation,
and it is also fundamental in understanding the quantum world for
the interpretation that the emergence of the classical world from
the quantum world can be seen as a decoherence process due to the
interaction between system and environment. From (\ref{free
evolution}), one can easily conclude the decoherence factor

\begin{figure}
\centerline{\scalebox{1.2}[1]{\includegraphics{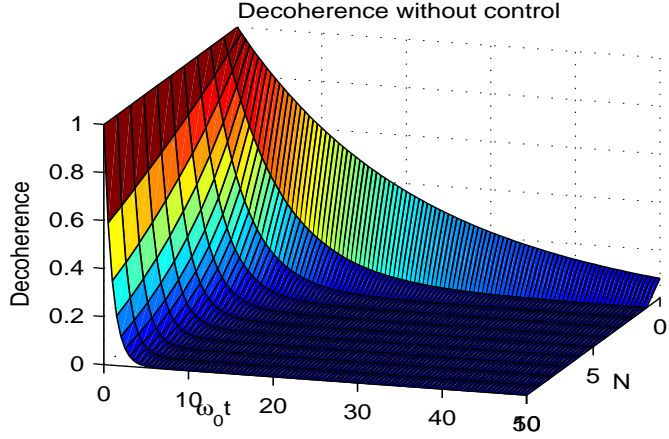}}}
\caption{(Color online)Time evolution of decoherence for
$N\in[0,10]$ without control.}
\end{figure}

\begin{figure}
\centerline{\scalebox{1.2}[1]{\includegraphics{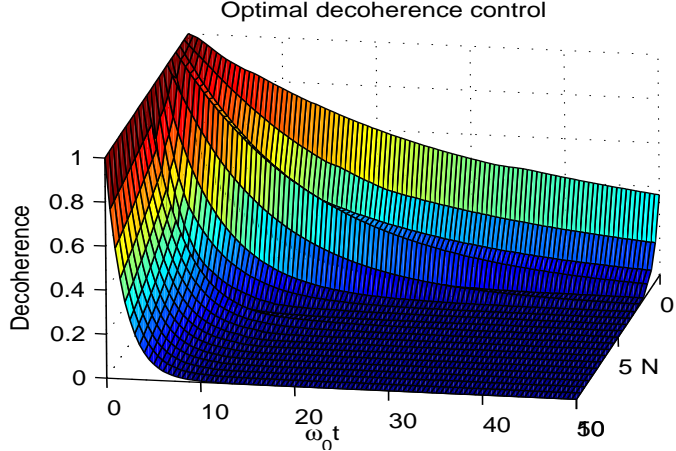}}}
\caption{(Color online)Time evolution of decoherence for
$N\in[0,10]$ with optimal control.}
\end{figure}

\begin{equation}
 \begin{array}{rcl}
\Lambda(t)&=&\frac{1}{2}{\sqrt{x_1^2(t)+x_2^2(t)}}\\
&=&\frac{1}{2}e^{-\frac{2N+1}{2}\gamma_0t}\sqrt{x_1^2(0)+x_2^2(0)}\\
&=&e^{-\frac{2N+1}{2}\gamma_0t}\Lambda(0),
 \end{array}
 \end{equation}
 decaying exponentially to $0$ with rate
 $\frac{2N+1}{2}\gamma_0$. In Figs.(9,10) we plot the time evolution of decoherence for continuous
 variation of
 $N$ (temperature).
Fig. (10) shows that at low temperature the decoherence can be well
controlled, i.e., the decoherence time can be delayed and its
amplitude can be amplified. With increasing $N$ (the temperature),
the environment induced fluctuations will be large enough to
neglecte the control field. Comparing with Fig. (9) we find that
these optimal controls can also prolong the decoherence time and
satisfy the controlled quantum logic gate operations, especially in
low temperature regime.

\section{Conclusions}
In the present work, we have studied the controlled population
transfer between selected quantum states in the Markovian open
quantum system, which can generate universal logic gates for quantum
computing. We use the optimal control method to control the
population transfer. Our numerical results indicate that the
occupation dynamics behaves differently for the different
environmental condition. The result can be summed up in three rules:
(i) The external control field plays an important role in population
transfer. (ii) The smaller the $N$ (temperature) the better the
control performance, especially when $N<0.2$ the completely
population transfer can be achieved. (iii) When temperature is high
enough the transfer is uncontrollable. Solid state qubits offer
remarkable advantages due to their scalability and controllability.
Therefore, optimal control of population transfer perhaps provides
an attractive approach to generate universal logic gates. We hope
such optimal control techniques to experimentally implement quantum
elementary logic gates in quantum computing networks in the near
future.

In this paper, we restrict the discussions to Markovian open quantum
systems and show the validity of our optimal control strategy to the
population transfer, and we analyze corresponding decoherence
behavior and we find that these controls can prolong the decoherence
time and satisfy the controlled quantum logic gate operation in low
temperature regime. To make the model more realistic in the
experiments and reduce the low temperature constraints it is worth
extending the scope to non-Markovian open quantum systems where the
seemingly lost information can return to the system at a later time,
which is our further work.

\section*{Acknowledgements}
We thank Prof. F. Nori for useful suggestions and enlightening
comments. This work was supported by the National Natural Science
Foundation of China (No. 60774099, No. 60821091), and the Chinese
Academy of Sciences (KJCX3-SYW-S01).



\begin{thebibliography}{10}
\bibitem{Schonfeldt:09}J. H. Sch\"{o}nfeldt, J. Twamley, S. Rebi\'{c}, Phys. Rev. A 80 (2009) 043401.


\bibitem{Stolze:08}J.~Stolze, D.~Suter, Quantum Computing: A Short Course from Theory to Experiment, Revised and Enlarged,
2nd Edition, Wiley, Berlin, Germany,
         2008.
\bibitem{Bellac:06}M. L.~Bellac, A Short Introduction to Quantum Information and Quantum
Computation,
 Cambridge University Press, Cambridge, UK, 2006.

\bibitem{Mermin:07}N. D.~Mermin, Quantum Computer Science: An
Introduction,
 Cornell University Press, New York, USA, 2007.

 \bibitem{Wiseman:93}
H. M. Wiseman, G. J. Milburn,  Phys. Rev. A 47 (1993) 642.
\bibitem{Wiseman:10}H. M.~Wiseman, G. J. Milburn, Quantum Measurement and
Control,
 Cambridge University Press, Cambridge, UK, 2010.

\bibitem{Rabitz:04}H. A. Rabitz,  M. M. Hsieh, C. M.
Rosenthal, Science 303 (2004) 1998.
\bibitem{Rabitz:10}C. Brif, R. Chakrabarti, H. Rabitz,  arXiv:0912.5121 (unpublished).
\bibitem{Balint:08}G. G. Balint-Kurti, S. Zou,  A. Brown, Adv.
Chem. Phys. 138 (2008) 43.

\bibitem{Facchi:05}
P. Facchi, S. Tasaki, S. Pascazio,  H. Nakazato, A. Tokuse, D. A.
Lidar, Phys. Rev. A 71 (2005) 022302.
\bibitem{Zhang1}J. Zhang, Y. X. Liu, F. Nori, Phys. Rev. A 79 (2009) 052102.
\bibitem{Dong:09}D. Dong, I. R. Petersen, New J. Phys. 11 (2009)  105033.
  \bibitem{Wu:06}R. B. Wu, T. J. Tarn, C. W. Li, Phys. Rev. A 73 (2006)  012719.

\bibitem{Rebentrost:09} P. Rebentrost, I. Serban, T.
Schulte-Herbrueggen, F.K. Wilhelm, Phys. Rev. Lett. 102 (2009)
090401.

\bibitem{Jirari:06} H. Jirari, W. P\"{o}tz , Phys. Rev. A 74  (2006)
022306.

\bibitem{Tarn:83}G. M. Huang,  T. J. Tarn, J. W. Clark, J. Math. Phys. 24 (1983) 2608.

\bibitem{Breuer:02}H.P.~Breuer, F.~Petruccione,The Theory of
         Open Quantum Systems, Oxford University Press, Oxford, UK,
         2002.

\bibitem{Wang:08}X. B. Wang, J. Q. You, F. Nori, Phys. Rev. A
77  (2008) 062339.

 \bibitem{Zhou:09}L. Zhou, S. Yang, Y. X. Liu, C. P. Sun, F. Nori, Phys. Rev. A
 80 (2009) 062109 .

 \bibitem{Cao:10}X. Cao, J.Q. You, H. Zheng, F. Nori, arXiv:1001.4831 (unpublished).

\bibitem{You:05}J. Q.~You, F. Nori, Physics Today
58, No. 11 (2005) 42.

\bibitem{Ashhab1}A. M. Zagoskin, S. Ashhab, J. R. Johansson, F. Nori, Phys. Rev. Lett.
97 (2006) 077001.

 \bibitem{Ashhab2}S. Ashhab, J.R. Johansson, F. Nori, Physica C
444  (2006) 45.

 \bibitem{Ashhab3}S. Ashhab, J.R. Johansson, F. Nori, New J. Phys.
 8 (2006) 103 .

 \bibitem{Neely:08}M. Neeley et al., Nature Physics 4  (2008) 523.

 \bibitem{Burgarth:09}D. Burgarth, K. Maruyama, M. Murphy, S. Montangero, T.
 Calarco, F. Nori,  M. B. Plenio, arXiv:0905.3373 (unpublished).

\bibitem{Burgarth:092}D. Burgarth, K. Maruyama, F. Nori, Phys. Rev. A 79
  (2009) 020305(R).


\bibitem{Liu:05}Y. X. Liu, J. Q.~You, L. F. Wei, C. P. Sun, F. Nori,
Phys. Rev. Lett. 95 (2005) 087001.

\bibitem{Deppe:08}F. Deppe, et al, Nature Physics
4  (2008) 686.

\bibitem{Wei:08}L. F.~Wei, J. R. Johansson, L. X. Cen, S. Ashhab, F. Nori,
 Phys. Rev. Lett. 100 (2008) 113601.


\bibitem{Weiss:99}U. Weiss, Quantum Dissipative System (2nd edition)
 World Scientific, London, UK,
         1999.

\bibitem{Table:06}M. Amniat-Talab,  R. Khoda-Bakhsh,  S. Gu\'{e}rinb,
Phys. Lett. A 359 (2006) 366 .
\bibitem{Makhlin:01}Y.~Makhlin, G. Sch\"{o}n, A. Shnirman, Nature(London) 386 (1999)
305.



\bibitem{Belavkin:83}V. P. Belavkin, Autom. Rem. Contr. 44 (1983) 178-188 .
\bibitem{Rice}S. A. Rice, M. Zhao, Optical Control of Molecular
Dynamics, Wiley, New York, USA, 2000.
\bibitem{Gordon}R. J. Gordon, S. A. Rice, Ann. Rev. Phys. Chem. 48 (1997) 601.

\bibitem{Brumer}P. Brumer and M. Shapiro, Principles of the
Quantum Control of Molecular Processes, Wiley Interscience, Hoboken,
NJ, 2003.

\bibitem{Alessandro}D. D'Alessandro, Introduction to Quantum Control and
Dynamics, Chapman \& Hall, Boca Raton, FL, 2007.








%
%

%
%
%
%
%
%

%
%
%
%
%
%
%
%
%
%
%
%
%
%
%
%
%
%
%
%
%
%
%
%
%
%
%
%
%
%
%
%
%
%
%
%
%
%
%
%


%
%
%

\end{thebibliography}
\end{document}